\documentclass[10pt,journal,compsoc]{IEEEtran}

\usepackage[left = 2cm, right=2cm, top=2cm, bottom=2cm]{geometry}
\usepackage{marginnote}
\usepackage{graphicx}
\usepackage{paralist}
\usepackage[inline]{enumitem}
\usepackage{mwe}
\usepackage{color,colortbl}
\usepackage{authblk}
\usepackage[]{hyperref}
\usepackage[]{amsmath}
\usepackage[]{amssymb, bm}
\usepackage{mathtools}
\usepackage{geometry}
\usepackage{pdflscape}
\usepackage{rotating}
\usepackage{array}
\usepackage{booktabs}
\usepackage{wasysym}
\usepackage[]{tikz}
\usepackage[]{pgfplots}
\usepackage{xspace}
\usepackage{pdflscape}
\usepackage[linguistics]{forest}
\usepackage{soul}
\usetikzlibrary{shapes}
\usepackage{subcaption}
\usepackage{tablefootnote}
\usepackage[hang,flushmargin]{footmisc}
\usepackage{caption}
\usetikzlibrary{arrows, arrows.meta, shapes, 3d, calc, fit, positioning,automata, shapes.multipart}
\usepackage{forest}
\usetikzlibrary{automata, decorations.text, positioning,shapes,arrows}
\graphicspath{{Figures/}}

\newlist{initem}{enumerate*}{1}
\setlist[initem]{itemsep = 1.125in, label=(\arabic*)}

\tikzstyle{myBlock}=[draw, fill=blue!20!white, minimum size=0.5in, node distance=1in, text width=1.5cm, align=center]
\tikzstyle{myPath} = [-{Latex[length=2mm,width=2mm]}, line width=0.2mm]

\title{On the Elements of Datasets for Cyber Physical Systems Security}

\author[1]{Ashraf Tantawy \thanks{This work is currently peer-reviewed. Please cite the latest preprint version on arXiv.}}
\affil[1]{School of Computer Science and Informatics, De Montfort University, Leicester, UK}

\begin{document}
	

\maketitle

\begin{abstract}
Datasets are essential to apply AI algorithms to Cyber Physical System (CPS) Security. Due to scarcity of real CPS datasets, researchers elected to generate their own datasets using either real or virtualized testbeds. However, unlike other AI domains, a CPS  is a complex system with many interfaces that determine its behavior. A dataset that comprises merely a collection of sensor measurements and network traffic may not be sufficient to develop resilient AI defensive or offensive agents. In this paper, we study the \emph{elements} of CPS security datasets required to capture the system behavior and interactions, and propose a dataset architecture that has the potential to enhance the performance of AI algorithms in securing cyber physical systems. The framework includes dataset elements, attack representation, and required dataset features. We compare existing datasets to the proposed architecture to identify the current limitations and discuss the future of CPS dataset generation using testbeds. 
\end{abstract}


\IEEEpeerreviewmaketitle

\section{Introduction}
A Cyber Physical System (CPS) is an integration of physical processes, computations, and networking \cite{lee2016introduction}. CPS applications are omnipresent, including process control systems, power generation and distribution, manufacturing, autonomous vehicles, transportation, and healthcare. A distinguishing feature of CPSs is that they are mission-critical, requiring high-level of resilience to failures. Over the last decade, CPS design shifted from proprietary hardware and software to open-source, standardized, hardware, software, and communications. This shift is motivated by the faster development cycle, lower development and maintenance cost, and inter-operability between systems offered by standard open-source solutions. Despite these advantageous, adopting standard open-source solutions increased the cyber attack surface of systems that were one day thought of as secure-by-design.

A significant amount of research has been recently devoted to secure this new generation of open CPS. With the proliferation of AI, particularly Deep Learning (DL), the application of AI algorithms for defensive CPS security grew both in academia and industry \cite{Sakhnini2020}. Modern DL algorithms require large training datasets to achieve good results. Large CPS datasets are currently far-from achievable for two key reasons. First, CPS datasets are quite unique and distinct from IT datasets, therefore utilizing available IT security datasets does not capture the dynamics of CPS. Second, large datasets available from real-life CPSs such as manufacturing facilities, transportation systems, or healthcare are not released for obvious security reasons. In response, researchers elected to generate their own CPS datasets. This is achieved by building either laboratory-scale physical testbeds or virtual testbeds \cite{Conti2021}. Some researchers release their datasets for public use. However, the released datasets are still with limited use as they lack key elements required for research use. To overcome these limitations, researchers opt for building their own testbeds to generate datasets according to their research needs. This trend is likely to continue, especially with the absence of any standards to describe the required dataset features and exchange format. Due to these facts, the academic research progress may be hindered significantly and the research may be limited only to entities that have access to large confidential datasets.

This paper dissects CPSs to develop the key elements that should be included in published CPS security datasets. The motivation is twofold:
\begin{initem}
\item rich datasets enable the development of more resilient AI models for CPS security, and
\item enable the reuse of published datasets by other research groups without a need to rebuild their own testbeds, eliminating the redundant work that slows down the research progress.
\end{initem}
The main contributions of the paper are:
\begin{initem}
\item dissection of the general CPS architecture and extraction of key dataset elements with illustrative scenarios,
\item study of CPS cyber attacks and their representation in the dataset, 
\item analysis of required dataset features to develop more resilient CPS security solutions,
\item discussion on testbeds as the main dataset generation mechanism and providing an eyesight to solve the scalability problem, and
\item survey existing datasets and their coverage for the proposed framework.
\end{initem}

\begin{figure}[tb!]
\centering
\begin{tikzpicture}[scale=0.7, transform shape]
\node [draw, rounded corners, fill=lightgray, minimum width=2cm, minimum height=1cm, align=center] at (2,2) (plant){Physical System};
\node [draw, fill=gray, minimum width=2cm, minimum height=1cm, below=1cm of plant, align=center] (cyber){Cyber System};
\draw (4,1) node[circle,minimum size=15pt,inner sep=0pt,draw] (sensor) {S};
\draw (0,1) node[circle,minimum size=15pt,inner sep=0pt,draw] (actuator) {A};
\node[inner sep=0pt] (ATTACKER1) at (5,1) {\includegraphics[width=0.05\textwidth]{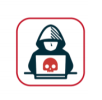}};
\node[inner sep=0pt] (ATTACKER2) at (-1,1) {\includegraphics[width=0.05\textwidth]{ATTACKER}};
\node[inner sep=0pt] (ATTACKER3) at (2,-1) {\includegraphics[width=0.05\textwidth]{ATTACKER}};
\draw [-stealth] (plant.east) -| (sensor.north);
\draw [-stealth] (sensor.south) |- (cyber.east);
\draw [-stealth] (cyber.west) -| (actuator.south);
\draw [-stealth] (actuator.north) |- (plant.west);
\end{tikzpicture}
\caption{CPS security dataset is influenced by the physical system, cyber system, attacker, and their interactions.}
\label{fig:CPS-ABSTRACT}
\end{figure}

Figure \ref{fig:CPS-ABSTRACT} is an abstract representation of the cyber physical system showing the three key components in the security context; the physical system, cyber system, and the attacker. The key elements of the dataset explained in this paper are extracted from the interactions between these three components. The paper is organized as follows: Section \ref{sec:CPS-ARCH} presents a general architecture for CPSs showing possible attack entry points. Sections \ref{sec:PHY-DATA}, \ref{sec:CYB-DATA}, and \ref{sec:ATTACK-DATA} explain physical data, cyber data, and attack data elements, respectively. Dataset attributes including labeling, class balancing, and scalability are discussed in Section \ref{sec:DATASET-FEATURES}. The role of testbeds in dataset generation is elucidated in Section \ref{sec:TESTBEDS}. Section \ref{sec:DISCUSSION} presents a discussion on the proposed framework and future research directions. Section \ref{sec:related-work} is a survey on existing datasets and their coverage for the proposed elements. The work is concluded in Section \ref{sec:CONCLUSION}.

\section{CPS Architecture: Attack Entry Points} \label{sec:CPS-ARCH}
\begin{figure}[tb!]
	\centering
	\includegraphics[scale=0.235]{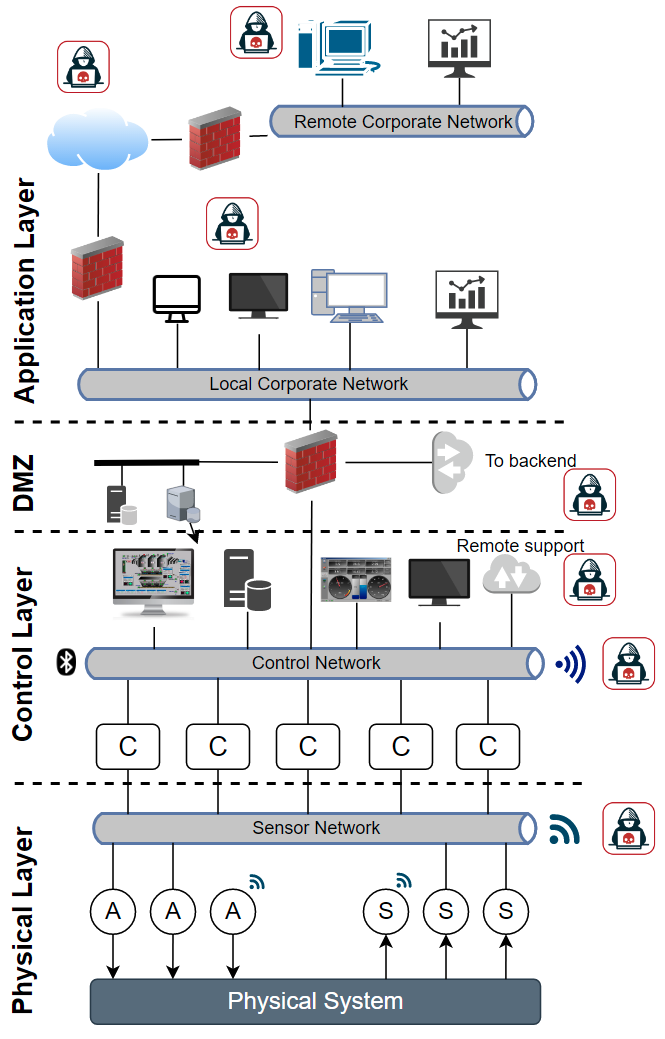}
	\caption{General CPS architecture. Entry points for remote cyber attackers are shown.}
	\label{fig:CPS-ARCH}
\end{figure}

Figure \ref{fig:CPS-ARCH} shows the general CPS architecture. The Physical layer comprises the physical system, sensors for system monitoring, and actuators for system control. A sensor network connects smart sensors and actuators that have communication capabilities. For a geographically dispersed CPS, e.g. a manufacturing facility, the sensor network is usually referred to as the Field network. For traditional hard-wired devices, sensors and actuators are connected in a point-to-point fashion to their respective controllers, and there is no sensor network. Examples of sensor networks include Fieldbus Foundation for the process control industry \cite{vincent2001foundation}, IEC 62026-3 (DeviceNet) for the manufacturing industry \cite{switchgear2000controlgear}, DNP3 for power systems \cite{east2009taxonomy}, SENT SAE-J2716 for the automotive industry \cite{costin2011j2716}, and Wireless Zigbee for home automation \cite{safaric2006zigbee}.

The Control layer comprises a set of controller nodes that utilize sensor measurements to control the physical system via actuators. The controllers are interconnected using a control network. Examples of control network protocols are CANbus for the automotive industry \cite{bozdal2018survey}, Modbus/TCP for the process industry \cite{bashendy2020design}, and DNP3 over TCP/IP for the power industry \cite{east2009taxonomy}. The control network typically has Human Machine Interface (HMI) nodes for monitoring and supervisory control. The HMI could be in the form of a dashboard as in the automotive industry, or an operator workstation as in the process control industry. Programming nodes may optionally be connected to the control network to modify and download controllers' software. In some industries, e.g., process control and power generation, control modifications are frequent, resulting in a permanent attachment of a programming machine to the control network. In other industries with infrequent control software update, e.g., automotive industry, the programming machine is attached to the control network as needed (typically during vehicle maintenance), or the software is downloaded Over The Air (OTA) from a remote server \cite{chowdhury2018safe}. The control network may be connected to the internet for remote technical support, although this practice is strongly discouraged. Current industry practice is to grant local/remote elevated access to the control network in very limited circumstances due to the associated security risk. The control network also may support wireless communication to connect with external wireless devices, such as Bluetooth devices in vehicles or RC Joysticks in UAVs. In such cases, the control network is typically segmented to secure the system.

Some variants of the control layer architecture may exist, depending on the CPS domain and the technology used. First, in distributed applications, the controller function could be embedded on the smart actuator, eliminating the need for a separate controller. In some industries, e.g., process control, stand-alone controllers such as a Programmable Logic Controller (PLC), are used as a backup to embedded field controllers implemented on the smart actuators. Second, the control layer may be composed of several hierarchical networks to support different levels of supervisory control and decision making such as the Supervisory Control and Data Acquisition System (SCADA). Finally, the control network may be segmented for enhanced security and organization of control functions, e.g., infotainment CAN and Drivetrain CAN in a vehicle.

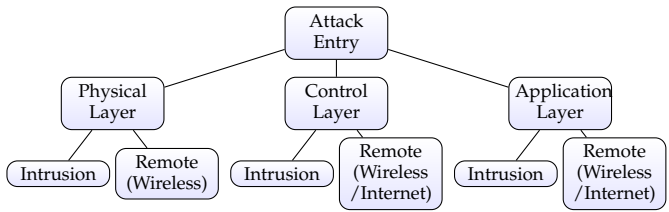
\begin{figure}[b]
	\centering
	\scriptsize

	\begin{tikzpicture}[scale=0.85, sibling distance=7em, every node/.style = {shape=rectangle, rounded corners, draw, align=center, top color=white, bottom color=blue!10}, text width=1.2cm]
		\tikzstyle{level 1}=[sibling distance=35mm, level distance=1.1cm]
		\tikzstyle{level 2}=[sibling distance=17mm, level distance=1.1cm]
		\tikzstyle{level 3}=[sibling distance=15mm]
		\tikzstyle{level 4}=[sibling distance=15mm]
		\node {Attack Entry}
		child
		{
			node {Physical Layer}
			child
			{
				node {Intrusion}
			}
			child
			{
				node{Remote (Wireless)}
			}
		}
		child
		{
			node {Control Layer}
			child
			{
				node {Intrusion}
			}
			child
			{
				node{Remote (Wireless /Internet)}
			}
		}
		child
		{
			node {Application Layer}
			child
			{
				node {Intrusion}
			}
			child
			{
				node{Remote (Wireless /Internet)}
			}
		};
	\end{tikzpicture}
	\caption{Attack entry methods for the CPS architecture.}
	\label{fig:ATTACK-ENTRY-METHODS}
\end{figure}

The application layer varies significantly across different CPS domains, and it could be further decomposed into several hierarchical layers. In general, the application layer reflects the business side of the CPS domain, and typically comprises a corporate network with a variety of computing nodes that perform data analytics on CPS-collected data to support business decision-making. The corporate network may be locally-located with the CPS or remotely-located, and in some architectures the corporate network is split into a local network and one or more remote networks. The process control and power industries typically have a hybrid architecture of both local and remote corporate networks, while the automotive/UAV industry has a remote backend that connects to vehicles. Two security mechanisms are typically used to avoid insecure direct communication between the application layer network and control layer network: A firewall that separates the two networks, and a DeMilitarized Zone (DMZ) that acts as a real-time data buffer between the two networks \cite{stouffer2011guide}. Figure \ref{fig:CPS-ARCH} shows also the attack entry points, which are crucial to understand the CPS attack surface and possible traces, hence the required dataset elements. The cyber system is reachable either by gaining physical access to the cyber network at any layer, or remotely over a communication link. Figure \ref{fig:ATTACK-ENTRY-METHODS} summarizes the attack entry methods for the general CPS architecture.

\section{Physical System Data} \label{sec:PHY-DATA}
\begin{figure}[tb!]
\centering
\begin{tikzpicture}[scale=0.7, transform shape]
\node [draw, rounded corners, fill=lightgray, minimum width=2cm, minimum height=1cm, align=center] at (2,2) (plant){Physical System \\ $\mathcal{M(\mathbf{x,u};\bm{\theta})}$};
\node [draw, fill=cyan, minimum width=2cm, minimum height=1cm, below=1cm of plant, align=center] (controller){Controllers \\ $\mathcal{A(\mathbf{y};\bm{\eta})}$}; 
\node[inner sep=0pt] (Human) at (4,-1) {\includegraphics[width=.05\textwidth, trim = 3 0 0 0, clip]{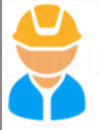}};
\draw [stealth-] (plant.west) -- ++ (-0.75,0) node[pos=1](input){} node[midway,above] {$\mathbf{u}$};
\draw [-stealth] (plant.east) -- ++ (1.5,0) node[midway](output){} node[midway,above]{$\mathbf{y}$};
\draw [-stealth] (output.center) |- (controller.east);
\draw [] (controller.west) -| (input.center);
\draw [stealth-] (plant.north) -- ++ (0, 0.75) node[midway,right]{$\mathbf{d}$};
\draw [<->](Human.west) -| (controller.south);
\end{tikzpicture}
\caption{Feedback control system. Human intervention via HMI nodes can alter/bypass the controller behavior.} 
\label{fig:FEEDBACK-SYSTEM}
\end{figure}
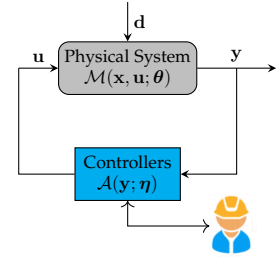

Figure \ref{fig:FEEDBACK-SYSTEM} illustrates the classical structure of a feedback control system. The physical system block integrates the physical system, sensors, and actuators in Figure \ref{fig:CPS-ARCH}. The system model is designated by $\mathcal{M}$. The input to the system is the action vector, $\mathbf{u}$, sent by the controllers to the system actuators, and the output from the system is the measurement vector, $\mathbf{y}$, representing sensor measurements. The internal state of the physical system is denoted by the state vector $\mathbf{x}$. The physical system is subject to an external disturbance vector, $\mathbf{d}$, that may alter the system behavior hence feedback control is required. The controller is a general term used to describe both regulatory as well as safety controllers that partially or fully shut down a system when entering a hazardous operating zone. The controllers are the nodes connected to both the sensor network (if applicable) and the control network in Figure \ref{fig:CPS-ARCH}. We designate the controller algorithm by $\mathcal{A}$. In real-life applications, the physical system is decomposed into several components, and each component is controlled by one or more controllers. Therefore, Figure \ref{fig:FEEDBACK-SYSTEM} could be interpreted as the feedback control system of a subsystem component as well. The human input to the controller represents human intervention via HMI nodes or similar on the control network in Figure \ref{fig:CPS-ARCH}. This human input may alter the control algorithm behavior or bypass it altogether. In the following, we study the importance of the elements of the tuple $\langle \mathbf{u,y,d}, \mathcal{M,A} \rangle$ for CPS security datasets.

\subsection{System Inputs and Outputs}
System input and output data ($\mathbf{u,y}$) enable us to use data-driven approaches, e.g., machine learning and system identification algorithms, to build a system model that captures the normal system behavior. This data is crucial for the CPS security dataset, and it is what mainly distinguishes CPS security from IT security datasets. This data has to be time-stamped and therefore synchronization between CPS nodes is essential. This type of data is the most common in CPS security datasets.

\subsection{System Disturbances}
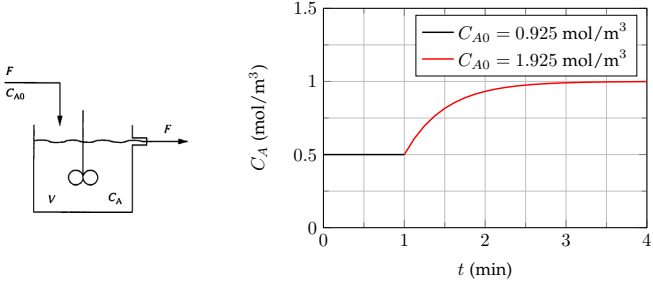
\begin{figure}
	\centering
	\begin{subfigure}{.2\columnwidth}
		\centering
		\includegraphics[scale=0.4]{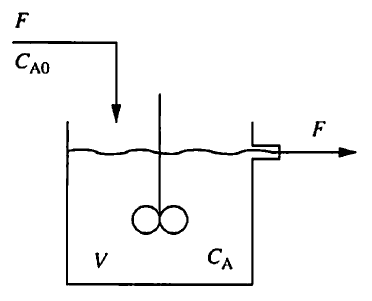}
	\end{subfigure}%
	\begin{subfigure}{1\columnwidth}
		\centering
		\begin{tikzpicture}[scale=0.7]
			\begin{axis}[xmin=0, xmax=4, ymin=0, ymax=1.5, xlabel={$t$ (min)}, ylabel={$C_A$ (mol/m$^3$)}, grid=both, minor tick num=1, width=0.9\columnwidth, height=0.67\columnwidth, legend cell align={left}, legend pos=north east]
				\addplot[domain=0:1, thick]{0.5};
				\addplot[domain=1:4, thick, red]{0.5+0.5*(1-e^(-(\x-1)/0.5)};
				\legend{
					$C_{A0} = 0.925 \text{ mol/m}^3$,
					$C_{A0} = 1.925 \text{ mol/m}^3$
				}
			\end{axis}
		\end{tikzpicture}
	\end{subfigure}
	\caption{Left: Continuous Stirred Tank Reactor (CSTR) process, Right: CSTR outlet concentration behavior for inlet concentration step change. Model values: $K_p=0.5, \tau=0.5 \text{ min}, (C_{A0})_{\text{init}} = 0.925 \text{ mol/m}^3$}
	\label{fig:CSTR-DIST}
\end{figure}

A continuous-time state space model for a given physical system could be expressed as:
\begin{align}
\dot{\mathbf{x}}(t) = \bm{f}(\mathbf{x}(t),\mathbf{u}(t);\bm{\theta}), \quad \mathbf{y}(t) = \bm{g}(\mathbf{x}(t),\mathbf{u}(t);\bm{\theta})
\end{align}
where $\bm{\theta}$ is the parameters vector. The system disturbances vector $\mathbf{d}$ is a subset of the parameters vector $\bm{\theta}$. Therefore, different values for the disturbances $\mathbf{d}$ give rise to a family of system models and consequently system behaviors. If the disturbances are not measured and reported in the dataset, then the underlying assumption is that the disturbances are constant throughout the dataset collection process. Therefore, the accuracy of any data-driven security algorithm that utilizes the dataset will be contingent on the closeness of the current plant disturbance values to the benchmark values. Any significant disturbance changes may lead to inaccuracies in the security algorithms, e.g., false alarms. To illustrate this point, consider the Continuous Stirred Tank chemical Reactor (CSTR) process in Figure \ref{fig:CSTR-DIST}. The system input is the inlet flow $F$, the system output is the outlet product concentration $C_A$, reactor volume $V$ is a model parameter, and the inlet product concentration $C_{A0}$ is the plant disturbance. The solution of the state space model for a first-order chemical reaction CSTR with rate $k$ could be expressed as \cite{marlin1995process}:
\begin{align}
C_A &= (C_A)_{\text{init}} + K_p \left[ C_{A0} - (C_{A0})_{\text{init}} \right] \left( 1 - e^{-t/\tau}\right)  \label{eq:CSTR} \\
K_p &= \frac{F}{F+Vk}, \qquad \tau = \frac{V}{F+Vk}
\end{align}

%

Figure \ref{fig:CSTR-DIST} plots the outlet concentration $C_A$ vs time after a disturbance step change in the inlet concentration $C_{A0}$. If the dataset does not report any information about $C_{A0}$, the indicated behavior may be interpreted by an intrusion detection algorithm as an anomalous behavior since the outlet concentration has changed without any change in the input flow $F$. This fact remains valid with the closed-loop system because the input/output system relationship is no longer valid regardless of the controller attempt to return the output to its setpoint value.

\noindent$\diamond$ Essential: \emph{A dataset should include all measurable disturbances as part of the physical system features, along with system inputs and outputs, i.e., $\langle \mathbf{u}, \mathbf{y}, \mathbf{d} \rangle$.}


\subsection{System Failure Modes} \label{subsec:systems-modes}
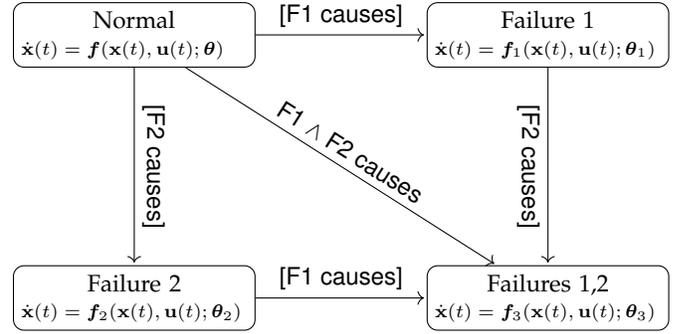
\begin{figure}[tb!]
	\centering
		\small
		\tikzset{every state/.append style={rectangle, rounded corners, text width = 3cm}}
		\begin{tikzpicture}[scale=0.8, shorten >=1pt,node distance=3.5cm,on grid,auto]
			\node[state] (N) at (0,0) {\centering{Normal \\ \scriptsize $\dot{\mathbf{x}}(t) = \bm{f}(\mathbf{x}(t),\mathbf{u}(t);\bm{\theta})$}}; 
			\node[state] (F1) [right=of N, xshift=2cm] {\centering{Failure 1 \\ \scriptsize $\dot{\mathbf{x}}(t) = \bm{f}_1(\mathbf{x}(t),\mathbf{u}(t);\bm{\theta}_1)$}}; 
			\node[state] (F2) [below=of N] {\centering{Failure 2 \\ \scriptsize $\dot{\mathbf{x}}(t) = \bm{f}_2(\mathbf{x}(t),\mathbf{u}(t);\bm{\theta}_2)$}};
			\node[state] (F12) [below=of F1] {\centering{Failures 1,2 \\ \scriptsize $\dot{\mathbf{x}}(t) = \bm{f}_3(\mathbf{x}(t),\mathbf{u}(t);\bm{\theta}_3)$}};

			\path[->] 
			(N) edge node [] {\textsf{[F1 causes]}} (F1)
			(N) edge node [sloped] {\textsf{[F2 causes]}} (F2)
			(F1) edge [sloped] node [below] {\textsf{[F2 causes]}} (F12)
			(F2) edge [] node {\textsf{[F1 causes]}} (F12)
			(N) edge [] node [sloped] {\textsf{F1 $\wedge$ F2 causes}} (F12);

	\end{tikzpicture}
	\caption{Modeling a CPS with known failure modes using a Hybrid automaton.}
	\label{fig:Hybrid-Automaton}
\end{figure}

Any CPS is subject to component failures during operation. When one or more components fail, the physical system model, whether physics-based or data-driven, is no longer valid, and the system behaves according to a new dynamical model. Therefore, anomaly detection systems that rely on such models may raise an attack alert whereas the system may just have a physical failure. Some common failure modes are known in advance. For such failures, a physics-based model could be developed if the physics of failure are understood. For parametric faults, the system model is given by the same original state space model with different parameter values:
\begin{align}
	\dot{\mathbf{x}}(t) &= \bm{f}(\mathbf{x}(t),\mathbf{u}(t);\bm{\check{\theta}})
\end{align}
where $\bm{\check{\theta}}$ is the parameter vector after failure. For structural faults, the system model is typically defined by a different state space model:
\begin{align}
	\dot{\mathbf{x}}(t) &= \bm{\check{f}}(\mathbf{x}(t),\mathbf{u}(t);\bm{\theta}')
\end{align}
If the physics of failure are not well-understood or difficult to model, then the failure scenario could be experimented in real life and the data collected could be used to develop a data-driven model for the failure mode. The dynamics of the normal and faulty system modes could be modeled in a unified framework using hybrid automata modeling approach \cite{henzinger2000theory}. Figure \ref{fig:Hybrid-Automaton} is an example hybrid automaton for a system with two types of faults and their combination. The dynamics of each discrete mode could be represented either by a state-space or a data-driven model. Machine learning algorithms could utilize the hybrid automaton model in the same way a healthy system model is used to distinguish between a cyber attack and system failure scenarios. However, it should be noted that attack scenarios that replicate the behavior of the hybrid automaton (though less-likely) cannot be distinguished from system failures using physical system data only.

The detection of unknown failure modes relies mainly on anomaly detection using unsupervised learning techniques \cite{purarjomandlangrudi2014data}. Using a variety of distance metrics, the system is considered in a failure mode if the distance between the current data and the nominal behavior benchmark is statistically significant. Since this approach is the same as anomaly detection for cyber attacks that relies only on physical system data, it may not be possible to distinguish between a cyber attack and a system failure. Therefore, for both known and unknown failure cases, it is a challenge to differentiate between a cyber attack and a system failure by relying on physical system data only. Cyber data can help to distinguish between these two scenarios since a cyber attack typically leaves a fingerprint in the cyber data, contrary to a system failure. This is another motivation for integrating physical and cyber data for multimodal cyber attack detection. Cyber data is discussed in details in section \ref{sec:CYB-DATA}.


In summary, it is strongly desirable to include all modes of operation of the system, including normal and failure modes, in the dataset to reduce the number of false alarms resulting from unknown modes. Dataset records associated with these modes have to be labeled accordingly. For system failures, the number of failure modes may increase exponentially with system scale, and an engineering judgment is needed to include the most probable failures. This information is typically included in reliability databooks for each CPS domain.

\vspace{2mm}\noindent $\diamond$ Essential: \emph{A dataset should include all normal modes of operation and the most probable failure modes for a given CPS.}

\subsection{System Model}
There are many scenarios where it is desired to augment a published dataset for research purposes. For instance, class imbalance is typical for cyber security datasets. Instead of using oversampling techniques for the minority class (predominantly attack data), it is more accurate to generate synthetic data for the system behavior under different attack scenarios. This is particularly important if the dataset is collected using a physical testbed and some attack scenarios may be infeasible to implement as they would cause a system hazard. As another example, in real systems, some system states and disturbances cannot be measured because they are either too expensive to be measured directly or because there is no supporting sensing technology. In all these scenarios, a system model $\mathcal{M}$ could be used to generate the required data or to estimate the required variables. As a simple example, consider the CSTR system behavior in (\ref{eq:CSTR}). If the inlet concentration disturbance variable is missing from the dataset, it is straightforward to show that it could be estimated by:
\begin{align}
	C_{A0} = (C_{A0})_{\text{init}} + \frac{C_A - (C_A)_{\text{init}}}{K_p \left( 1 - e^{-t/\tau}\right)}
\end{align}
which could be calculated at any given time provided that the inlet flow $F$, outlet concentration $C_A$, and initial inlet concentration are available. 

\noindent $\diamond$ Optional: \emph{A system physics-based model could augment a dataset to estimate unknown disturbances or generate a synthetic dataset.}

\subsection{Control Algorithms}
Assuming an automatic mode of operation, the system control algorithms impose additional constraints on input-output relationships. Together with the system model, a system trace on the following form could be generated by iteratively going around the closed loop system in Figure \ref{fig:FEEDBACK-SYSTEM}:
\begin{align}
\mathbf{u}[1] \xrightarrow[]{\mathcal{M}} \mathbf{y}[1] \xrightarrow[]{\mathcal{A}} \mathbf{u}[2] \xrightarrow[]{\mathcal{M}} \mathbf{y}[2] \ldots \xrightarrow[]{\mathcal{A}} \mathbf{u}[n] \xrightarrow[]{\mathcal{M}} \mathbf{y}[n] 
\end{align}

The idea of trace generation could be used for data preprocessing and model-based detection. Table \ref{tab:Control-Alg} shows the four cases where the system trace values satisfy/violate the system model $\mathcal{M}$ and/or the controller algorithm $\mathcal{A}$. Satisfaction of both $\mathcal{M}$ and $\mathcal{A}$ indicate a normal system operation with high probability. There is still a possibility that a stealthy attack may produce a valid system trace, but this requires either perfect system and control knowledge, or the random generation of coincidental valid trace, both have very low probability. Trace data that satisfy $\mathcal{A}$ but violate system model $\mathcal{M}$ may indicate a system failure. Violation of controller algorithm $\mathcal{A}$ is a strong indication of a cyber attack assuming very low probability of controller hardware/software bugs that could produce incorrect results. It should be noted that for the control algorithm to be useful in data tracing, the controller has to be in auto mode to ensure the controller is not bypassed. This information can be obtained from timed controller log. In addition, the parameter values of the controller have to be known. As an example, for a basic PID controller, proportional, derivative, integral constants and the setpoint have to be defined.

\noindent $\diamond$ Optional: \emph{A controller model could augment a dataset to assist in data preprocessing, labeling, and model-based detection methods.}

\begin{table}[tb!]
	\caption{\small System trace violation of physical model $\mathcal{M}$ or control algorithm model $\mathcal{A}$ could be utilized in data preprocessing and model-based detection.}
	\centering
	\begin{tabular}{c cc}
		\hline\hline
		 & $\mathcal{M}(\mathbf{u})=\mathbf{y}$ & $\mathcal{M}(\mathbf{u}) \neq \mathbf{y}$ \\ \hline
		$\mathcal{A}(\mathbf{y})=\mathbf{u}$  & Normal & Failure  \\	
		$\mathcal{A}(\mathbf{y}) \neq \mathbf{u}$ 	& Attack & Attack, Failure \\	\hline
	\end{tabular}
	\label{tab:Control-Alg}
\end{table}

\section{Cyber System Data} \label{sec:CYB-DATA}
Apart from physical and side channel attacks on sensors and actuators, all attacks on the CPS are injected via the cyber system. Therefore, looking into the physical system data in isolation does not give the full picture. Physical system input or output changes are the last step in a sequence of attack steps that may span from hours to months, leaving a trace in the cyber system, being in a host node or in network traffic. Hence, it is intuitive that both physical and cyber data should be jointly investigated. From Figure \ref{fig:CPS-ARCH}, cyber nodes can be categorized into computing nodes and communication nodes. Computing nodes include sensors, actuators, controllers, HMI nodes, log servers, and corporate workstations. Communication nodes are networking devices that form the sensor network, control network(s), and the backend corporate network. In this section, we investigate the elements of CPS datasets as related to these cyber nodes.


\subsection{Network Traffic Log}
Network traffic data is the most common cyber data in all available datasets. Ideally, network traffic data should be collected from all available sensor and control networks. If the time window for data collection is long, this will give rise to a massive amount of data. This data may have a lot of redundancy during stable CPS operation. This is because CPS traffic is near-deterministic (although the degree of determinism varies across CPS domains), as sensors, controllers, and actuators communicate periodically to exchange measurements and control actions. Therefore, it is important to pre-process the data as it is collected to remove any redundancies. This will reduce the dataset size, enhance the learning rate and reduce overfitting for AI algorithms, and minimize the class imbalance problem (discussed in Section \ref{subsec:class-imbalance}). Apart from redundant data, there are few operational events that result in a temporary irregular traffic. For example, a disturbance beyond controller operating range may require an operator intervention to switch the control mode to manual. This is a typical scenario in the process control industry and more recently in autonomous systems. A system hazardous state may trigger a safety action to prevent incidents. An operator may manually poll some information from a controller for monitoring purposes. In these scenarios, the intervention, whether human or automatic, introduces irregular traffic between different system nodes. It is important that the dataset captures these irregular traffic cases. Otherwise, defensive AI models will not be robust against irregular events. Manual injection of these actions during dataset generation is a laborious task, and automation is needed.


Communication protocols for CPSs vary significantly by domain. In addition, a single node may support multiple protocols to communicate with other nodes at different levels. For example, a controller may support different protocols to communicate with smart sensors (e.g., WirelessHART), HMI (e.g., Modbus TCP), and a log server (e.g., OPC). This multiplexing of protocols on a single node induces a specific timing behavior that may not be available with a single protocol per node. Therefore, it is important to embody this protocol diversity in the dataset to be representative of the given CPS domain.


\vspace{2mm}\noindent $\diamond$ Essential: \emph{A dataset should include a network traffic log for all available sensor and control networks. Traffic log should include all supported communication protocols in addition to regular and irregular traffic scenarios.}

\subsection{Host Log}
A host is any computing node, including smart sensors and actuators, controllers, HMI nodes, and servers (we exclude network equipment and their vulnerabilities from the discussion). The compromise of a host node requires a series of attack steps that typically involve a reconnaissance phase, communication with the remote node, and vulnerability exploitation to gain a node access, possibly followed by a privilege escalation. Most of these attack steps could leave multiple trace evidences in the host log, which could lead to attack discovery when combined with other network and physical features. Host logs are entirely overlooked in all available datasets.

\begin{figure}[tb!]
\centering
\begin{tikzpicture}[scale=0.7, transform shape]
\fontfamily{sffamily}{\fontsize{10}{10}\selectfont
\draw [thick] (0,0) -- (7,0);
\node[inner sep=0pt] (MITM) at (2,-1.5) {\includegraphics[width=.125\textwidth]{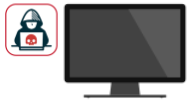}};
\draw [] (MITM.north) -- ($(0,0)!(MITM.north)!(6,0)$);
\node[inner sep=0pt] (HMI) at (1,1) {\includegraphics[width=0.125\textwidth]{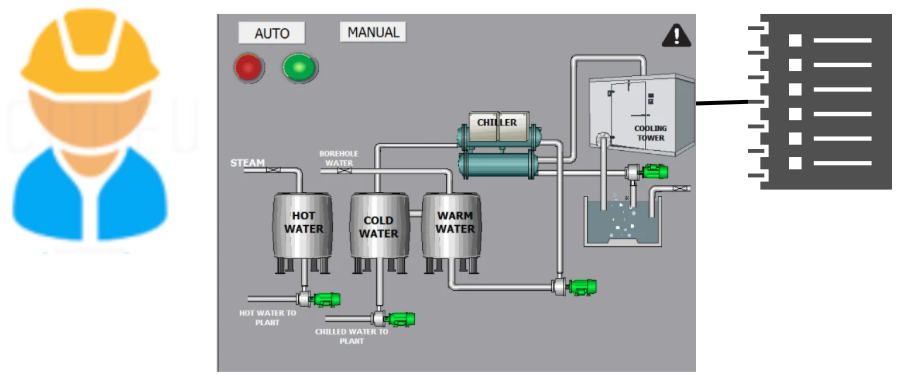}};
\draw [] (HMI.south) -- ($(0,0)!(HMI.south)!(6,0)$);
\node[draw, thick, fit={(5,-1.5) (7,-0.75)}, inner sep=0pt, label=center:Controller, rounded corners] (CONT) {};
\draw [] (CONT.north) -- ($(0,0)!(CONT.north)!(6,0)$);
\node[inner sep=0pt] (MOTOR) at (6,-2.5) {\includegraphics[width=0.1\textwidth]{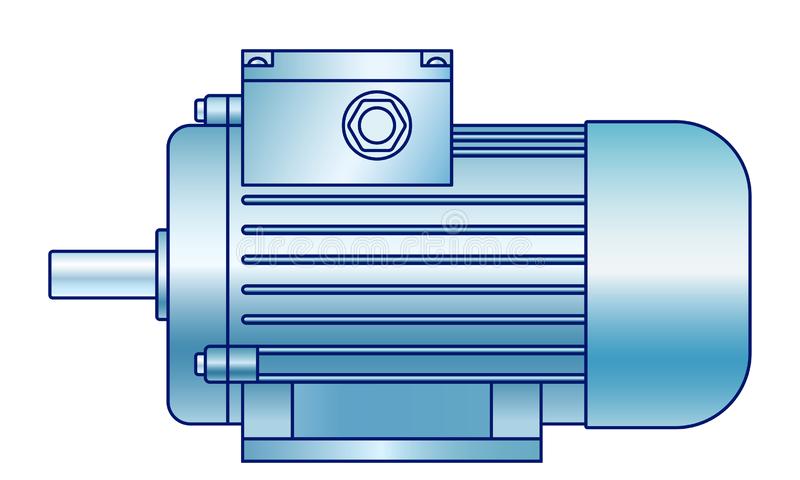}};
\draw [] (CONT.south) -- (MOTOR.north);
\node[] at (2.8,1.25) {\small Host log};
\node[] at (2.5, -2.5) {\small IP Spoofing};
\draw [->, dashed, rounded corners] (1.5, 0.5) -- (1.5,0.15) -- (5.75, 0.15) -- (5.75, -0.45);
\node[inner sep=0pt] (PACKET-HMI) at (3.5,0.35) {\includegraphics[width=0.15\textwidth]{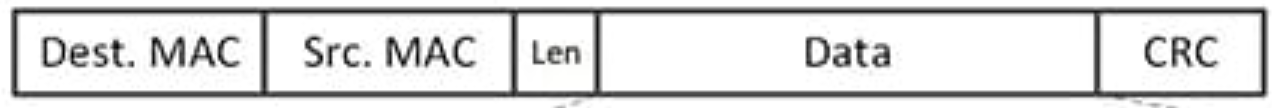}};
\draw [->, dashed, rounded corners] (2.25, -0.75) -- (2.25, -0.15) -- (5.4, -0.15) -- (5.4, -0.45);
\node[inner sep=0pt] (PACKET-MITM) at (3.75, -0.4) {\includegraphics[width=0.15\textwidth]{PACKET}};
}
\end{tikzpicture}
\caption{A host log may help differentiating between a legitimate operator action and an attacker compromise.}
\label{fig:HostLog-S1}
\end{figure}

A host log could be divided into Operating System (OS) logs and application logs. An OS log typically includes a variety of time-stamped events, such as running processes, system calls, authentication logs, file transactions, user actions, and system errors. Application logs vary significantly based on the application design and its security features. For example, the log for an HMI application that supports multiple user profiles may include time-stamped login data and various user action statistics. To illustrate the potential benefits of using host logs, consider the scenario in Figure \ref{fig:HostLog-S1}, where the controller is switched to manual mode followed by a command to increase the speed of rotation of a motor. This scenario could be initiated by an operator via the HMI, an attacker who compromised the HMI, or by another node via IP Spoofing attack that generates identical network traffic to the controller. For these three scenarios, the network and physical data are identical (ignoring the initial network fingerprint of the IP spoofing attack that fades with time). The host log data is what could make the distinction. For the operator case, the HMI host log data contains operator actions to override the controller. The compromised HMI case has this same data with the added HMI node compromise events. The IP spoofing attack case does not have any fingerprint in the host log. Table \ref{tab:IDS-DECISION} shows the legitimate and attack scenarios, versus three designs of the Intrusion Detection System (IDS); an IDS trained on physical data only, an IDS trained on physical and network data, and an IDS trained on physical, network, and host log data. For the first two IDS designs, the IDS is trained on this pattern using physical and network data, so the decision will always be "Negative", as it is oblivious to the initiator of the network traffic. The IDS trained on the augmented dataset with host log data is able to differentiate between the legitimate operator action and the HMI compromise/IP spoofing attacks, with the added advantage of attack classification.

\begin{table}[b!]
\caption{\small IDS decision for an operation scenario where the controller is bypassed and a control command is sent directly to the actuator. TN: True Negative, FN: False Negative, TP: True Positive.}
\centering
\begin{tabular}{c ccc}
\hline\hline
True State & \multicolumn{3}{c}{IDS Decision for Different Training Data} \\ [0.5ex]
\cline{2-4}
 & Physical & PHY \& Network & PHY \& NET \& Host \\
\hline 
Normal & TN & TN & TN \\
Attack & FN & FN & TP \\ 
\hline
\end{tabular}
\label{tab:IDS-DECISION}
\end{table}

\begin{figure}[b!]
\centering
\begin{tikzpicture}[scale=0.7, transform shape]
\fontfamily{sffamily}{\fontsize{10}{10}\selectfont
\draw [thick] (0,0) -- (7,0);
\node[inner sep=0pt] (MITM) at (2,-1.5) {\includegraphics[width=.125\textwidth]{MITM}};
\draw [] (MITM.north) -- ($(0,0)!(MITM.north)!(6,0)$);
\node[inner sep=0pt] (HMI) at (1,1) {\includegraphics[width=0.125\textwidth]{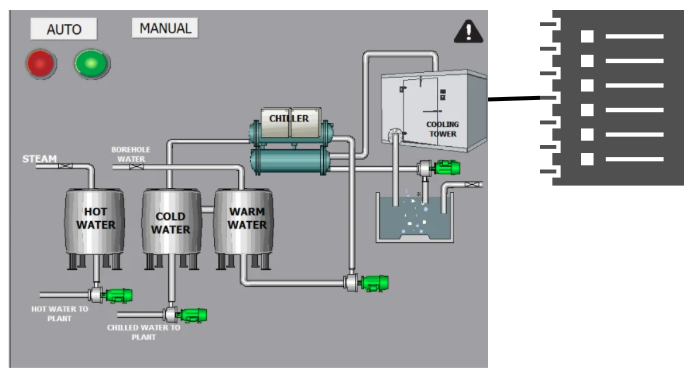}};
\draw [] (HMI.south) -- ($(0,0)!(HMI.south)!(6,0)$);
\node[draw, thick, fit={(5,-1.5) (7,-0.75)}, inner sep=0pt, label=center:Controller, rounded corners] (CONT) {};
\node[inner sep=0pt] (CONT-LOG) at (7.5,-1.125) {\includegraphics[scale=0.2]{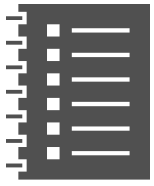}};
\draw [] (CONT.east) -- (CONT-LOG.west);
\draw [] (CONT.north) -- ($(0,0)!(CONT.north)!(6,0)$);
\node[inner sep=0pt] (MOTOR) at (6,-2.5) {\includegraphics[width=0.1\textwidth]{motor}};
\draw [] (CONT.south) -- (MOTOR.north);
\node[] at (2.8,1.25) {\small Host log};
\node[] at (2.5, -2.5) {\small MITM};
\draw [->, dashed, rounded corners] (1.75, -0.75) -- (1.75,-0.25) -- (0.5, -0.25) -- (0.5,0.25);
\node[inner sep=0pt] (PACKET-HMI) at (.25,-0.5) {\includegraphics[width=0.15\textwidth]{PACKET}};
\node [] at (0.25, -0.75) {\scriptsize Speed = 500 rpm};
\draw [->, dashed, rounded corners] (2.25, -0.75) -- (2.25, -0.15) -- (5.4, -0.15) -- (5.4, -0.45);
\node[inner sep=0pt] (PACKET-MITM) at (3.75, -0.4) {\includegraphics[width=0.15\textwidth]{PACKET}};
\node [] at (3.75, -0.75) {\scriptsize Speed = 1000 rpm};
}
\end{tikzpicture}
\caption{Host log and the detection of stealthy attacks.}
\label{fig:HostLog-S2}
\end{figure}

The host log may also enable the discovery of stealthy attacks. In a stealthy attack, the attacker manipulates local controller data while transmitting fake normal data over the network to evade the receiving nodes. This scenario is depicted in Figure \ref{fig:HostLog-S2}, where an attacker node acting as a Man In The Middle (MITM) node sends fake normal data to the HMI while sending hazardous data to the controller. The controller log will include the malicious data if the attacker was not able to compromise the controller to overwrite the log. In this case, HMI and controller logs will not match. Alternatively, if the attacker compromised the controller, controller log may have an attack fingerprint. In all possible cases, a host log represents an additional source of information that makes it harder for the attacker to make different information sources consistent, hence increases the probability of attack detection. Finally, on the local node level, the host log has a wealth of information to support the development of a probabilistic model (profile) for each node/node user. This enables an early detection of attacks on the node level before the attack spreads to other network nodes. On the system level, the fusion of host logs from system nodes could reveal the sequence of attack steps across nodes, e.g. pivot attacks. These patterns could be discovered by AI algorithms only if the host logs are available and aggregated.

\vspace{2mm}\noindent $\diamond$ Essential: \emph{A dataset should include a host log for each node, including the controller and HMI nodes as a minimum.}

\section{Cyber Physical System Attacks} \label{sec:ATTACK-DATA}
Attacks could be classified according to the security objective into Confidentiality, Integrity, and Availability (CIA) attacks. In CPS, confidentiality is arguably not the main attack objective \emph{at the control network level}, but rather the system safety. Therefore, we limit our discussion to integrity attacks where data is manipulated, and availability attacks where the objective is the denial of system services. One important characteristic of a realistic dataset is to include labeled data for the whole attack process, starting from the reconnaissance stage until reaching the final attack goal. In this section, we summarize the attack types that need to be included in realistic datasets.

\subsection{Integrity Attacks}
\begin{figure}[tb!]
	\centering
	\begin{tikzpicture}[scale=0.7, transform shape]
		\node [draw, rounded corners, fill=lightgray, minimum width=2cm, minimum height=1cm, align=center] at (2,2) (plant){Physical System};
		\node [draw, circle] at (4.5,1) (sensor){S};
		\node [draw, fill=cyan, minimum width=2cm, minimum height=1cm, below=1cm of plant, align=center] (controller){Controller};
		\node at (5.5,1.75) (inj1){1};
		\draw [-> , rounded corners, red] (5.4,1.75) -- +(-0.4,-0.25) -- +(0,-0.25) -- +(-0.4,-0.5);
		\node at (-1.5,1.75) (inj3){3};
		\draw [-> , rounded corners, red] (-1.4,1.75) -- +(0.4,-0.25) -- +(0,-0.25) -- +(0.4,-0.5);
		\node at (1.25,-1) (inj2){2};
		\draw [-> , rounded corners, red] (1.35,-1) -- +(0.4,0.25) -- +(0,0.25) -- +(0.4,.5);
		\node [draw, circle] at (-0.5,1) (actuator){A};
		\node[inner sep=0pt] (HMI) at (2,-1.75) {\includegraphics[width=.05\textwidth]{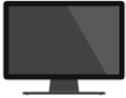}};
		\node at (2.75,-1.25) (DMLabel){DM};
		\draw [stealth-, dashed] (HMI.east) -- (5.5,-1.75) |- (sensor.east);
		\draw [-stealth, dashed] (HMI.west) -- (-1.5,-1.75) |- (actuator.west);
		\draw [>=triangle 45, <->] (HMI.north) -- (controller.south);
		\draw [-stealth] (plant.east) -| (sensor.north);
		\draw [-stealth] (sensor.south) |- (controller.east);
		\draw [-stealth] (controller.west) -| (actuator.south);
		\draw [-stealth] (actuator.north) |- (plant.west);
	\end{tikzpicture}
	\caption{Integrity attack injection points. An HMI is shown as an example DM node. Dashed lines are for smart sensors/actuators.} 
	\label{fig:INT-ATTACK}
\end{figure}
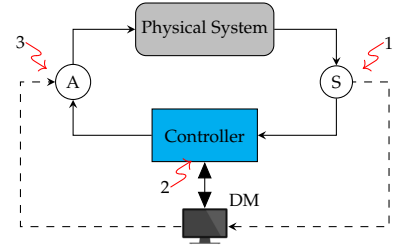

In integrity attacks, the attacker tampers with the data streams that carry system measurements and/or control actions. This could be at the end node post compromise or by manipulating network traffic through the data stream. Three data types that can disrupt system operation are false data, late data, or missing data. Network traffic manipulation is done by intercepting a packet then either dropping it (missing data), submitting it later (replay attack, late data), or modifying and resubmitting it (false data). Figure \ref{fig:INT-ATTACK} shows the key data streams, which are sensor $\rightarrow$ controller $\rightarrow$ actuator, sensor/controller $\rightarrow$ DM, and DM $\rightarrow$ controller/actuator, where DM designates higher-level Decision-Making nodes beyond the low-level controller, e.g., HMI or a supervisory/safety controller. The DM-controller data stream is for centralized architectures where sensors/actuators are hardwired to the controller, while sensor-DM and DM-actuator data streams are for smart sensor/actuator nodes. Although tampering could be random, often times the objective is to cause the physical system to move to or remain in an unsafe state and to prevent any automatic or manual corrective action. This is done by manipulating actuator commands either directly or indirectly through the sensor $\rightarrow$ controller $\rightarrow$ actuator stream. The shortest path is to manipulate the actuator directly (Figure \ref{fig:INT-ATTACK}, injection point 3). However, this may not always be possible depending on the CPS architecture and available vulnerabilities. A longer path is to manipulate sensor data or controller parameters, such as a setpoint or tuning parameters (Figure \ref{fig:INT-ATTACK}, injection point 2 and 3, respectively). It should be noted that for hardwired centralized systems, a compromise of the central controller enables the attacker to manipulate sensor, actuator, and controller data alike. This is a clear drawback of centralized architectures that are still quite common in different CPS domains.

A key question for integrity attacks is \emph{how} the data will be manipulated. To define the signal waveform to be injected, several factors come into play; the system dynamics, current system state, the speed by which the attacker wants to reach the unsafe system state, and whether a stealthy attack is required. There is no single waveform that fits all systems, and the \emph{optimal} waveform depends on the attack objective, system model, as well as existing detection methods. These factors have to be taken into account when generating attacks for datasets. Unfortunately, the majority of existing datasets utilize classical waveforms such as the step, pulse, and ramp functions, independent of the given system. The design of injection attack waveforms for a given system and attack requirements is important for both design-time verification and run-time penetration testing, yet is not explored fully in the literature (an exception is the control-theoretic approach to stealthy attacks, as discussed in the next section). With the lack of analytical or algorithmic methods to identify the optimal waveforms, experiments on a simulated system model may be a viable alternative.

\subsection{Stealthy Attacks}
The notion of a \emph{stealthy attack} has a slightly different meaning in different research communities. In control-theoretic research, it is assumed that the injected values are transmitted as-is to the decision-making nodes. Therefore, the main research focus is on designing the injected waveforms to avoid detection \cite{mo2015performance}. By contrast, in CPS security research, the injected data and the transmitted data to the decision-making nodes are decoupled, and the attacker can inject whatever data values that achieve the attack objective while transmitting fake data to the decision-making nodes. The stealthy signal approach is more applicable to CPS domains where a sudden state change could be observed even if malicious data is sent to the observer, e.g., aircrafts, while sensor $\rightarrow$ DM data stream injection approach is applicable to CPSs that are not directly-observable, e.g., process control plants.

Stealthy attacks launched by tampering with the sensor $\rightarrow$ DM data stream could result in dangerous scenarios. Table \ref{tab:stealthy-attacks} summarizes the four scenarios based on the current system state and whether the sensor $\rightarrow$ DM stream is tampered with. Tampering here refers to reporting the opposite system state to the DM (i.e., reporting safe measurements when the system is in an unsafe state and vice versa). In the first row in Table \ref{tab:stealthy-attacks}, when the system is unsafe (whether naturally or driven by an integrity attack on the sensor $\rightarrow$ controller $\rightarrow$ actuator path), the true information is reported which results in an intervention to safeguard the system. In the second row, when the system is in a safe state and malicious unsafe measurement information is reported, An unnecessary intervention will be triggered that may result in a system shutdown. Although this type of attack may result only in system disruption and financial losses, it is easier than driving the system into an unsafe state (see a more detailed discussion in \cite{tantawy2021automated}). The bottom right cell in Table \ref{tab:stealthy-attacks} is the most dangerous, where the system is in an unsafe state, while the information reported to decision-making nodes reflect a safe system state. This will prevent any protective or corrective action for the system, and the consequences could be catastrophic. In general, the design of a stealthy attack is not a trivial task, as real world systems are much more complicated than toy systems treated in the literature. A typical CPS may have hundreds of components coupled together, and an attack on one component will manifest itself in its connected components. The design of a stealthy attack in such case would require the manipulation of all impacted process variables.


\begin{table}[tb!]
\centering
\begin{tabular}{p{2cm}p{3.25cm}p{2.25cm}} 
 \hline
 Sensor-DM Stream & Safe State & Unsafe State (Natural/Attack) \\ [0.5ex] 
 \hline\hline
 True State & Normal Operation & DM Intervention \\ 
 Malicious State & DM Intervention (futile) & Hazard \\
 \hline
\end{tabular}
\caption{A stealthy attack by manipulating sensor-DM stream results in a futile intervention or a system hazard.}
\label{tab:stealthy-attacks}
\end{table}

A CPS security dataset has to include the three combinations of integrity attacks in Table \ref{tab:stealthy-attacks}. For stealthy attacks, the sensor data logged in the dataset must be the malicious data, and not the true physical system measurements, as the malicious data is the actual information sent to DM nodes during the attack. Depending on the number and location of DM nodes for the given CPS, the dataset may include \emph{partial} stealthy attacks, where malicious information is sent to some, but not all, DM nodes. This is an easier case for the intrusion detection system than the complete stealthy attack case, yet includes likely variations in real-life scenarios. Currently, none of the available datasets has such a level of sophisticated attacks and variations, and therefore, detection of injected attacks in the datasets does not represent a major challenge. Incorporation of such attacks could promote new types of intrusion detection systems, e.g., multi-modal detectors that utilize data from different sources along with system models as discussed in this work.

\vspace{2mm}\noindent $\diamond$ Essential: \emph{A dataset should include integrity attacks, including variants of stealthy attacks with logged malicious data.}

\subsection{Denial of Service (DoS) Attacks}
DoS attacks aim to shutdown the services provided by the cyber system. For example, a system controller has the main function of regulating the actuators based on sensor measurements and the control algorithm, and a secondary function of reporting the data to the HMI and log servers. A DoS attack against a controller aims to shutdown both control and reporting services, typically by flooding the controller with a large number of packets in a short time to consume the controller resources. There are two key points for DoS attacks. First, the impact of the DoS attack depends on the cyber system fail-safe configuration and the physical system/environment state. As an example, if the cyber system is configured to use the \emph{last-known-good-value} for actuator outputs, then the DoS attack will not have an impact if the system is stable with no disturbances. Second, most controllers deployed in a mission-critical CPS run one form or another of a real-time operating system (RTOS), which deploys an RT scheduling algorithm that gives a high priority for control tasks regardless of other conditions. Therefore, a DoS attack may have an insignificant impact on the control services and system stability. These facts are overlooked in existing datasets that include DoS attacks. Datasets with DoS attacks require logging labeled network, physical, and host data to assess the real impact of the attack and to design the appropriate detection and countermeasure algorithms.


\vspace{2mm}\noindent $\diamond$ Essential: \emph{A dataset should include DoS attacks with/without synchronized system disturbances and related host log data for actual scheduling times for control and communication tasks.}

\subsection{Zero-day Attacks}
\begin{figure}[tb!]
	\footnotesize
	\centering
	\begin{tikzpicture}[scale=0.8, sibling distance=7em, every node/.style = {shape=rectangle, rounded corners, draw, align=center, top color=white, bottom color=blue!10}]
		\tikzstyle{level 1}=[sibling distance=25mm, level distance=1.0cm]
		\tikzstyle{level 2}=[sibling distance=45mm, level distance=0.9cm]
		\tikzstyle{level 3}=[sibling distance=15mm]
		\tikzstyle{level 4}=[sibling distance=15mm]
		\node {Attack}
		child { node {Availability (DoS)} }
		child
		{
			node {Integrity}
			child
			{
				node {S $\rightarrow$ C $\rightarrow$ A}
				child
				{ 
					node {MITM}
					child {	node {Drop}}
					child { node{Replay}}
					child { node{Modify}}
				}
				child { node {Host} }
			}
			child 
			{ 
				node {DM}
				child
				{ 
					node {MITM}
					child {	node {Drop}}
					child { node{Replay}}
					child { node{Modify}}
				}
				child { node {Host} }
			}
		}
		child {node {Zero-day} };
	\end{tikzpicture}
	\caption{Essential CPS attacks for a dataset. S $\rightarrow$ C $\rightarrow$ A stands for Sensor $\rightarrow$ Controller $\rightarrow$ Actuator path.} 
	\label{fig:AttackTaxonomy}
\end{figure}
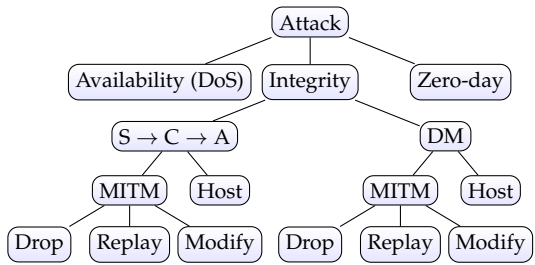

Zero-day attacks cannot be included in the training dataset because by definition they exploit new system vulnerabilities. However, they can be included in the test dataset to evaluate the performance of detection and response systems to unseen attacks. Zero-day attack detection relies mainly on anomaly detection methods that build a baseline model \cite{omar2013machine}. Therefore, zero-day attacks should not induce data records significantly deviating from the baseline, otherwise their detection will be a trivial task. Figure \ref{fig:AttackTaxonomy} summarizes the types of attacks discussed in this section.

\vspace{2mm}\noindent $\diamond$ Essential: \emph{A test dataset should include attacks not present in the training dataset, which are not significantly deviating from normal system behavior.}

\subsection{Attack Vectors}
Existing datasets assume the presence of the attacker on the control network. This assumption is valid only in case of insider attacks. In real attack scenarios, the attacker reaches the control network after several attack steps, starting from an entry point such as a wireless access point or a remote internet connection. These steps leave a footprint on different logs, including host, network, and physical system logs. A realistic dataset needs to include full attack vectors. The challenge here is that there could be many attack entry points, and the different combinations of attack vectors may be prohibitively large. Some heuristics are needed, such as selecting the most probable/vulnerable attack entry points, the elimination of attacks that have the same or similar fingerprint on the system, and grouping similar nodes according to a given criterion, e.g., OS, running services, and communication protocols.

\vspace{2mm}\noindent $\diamond$ Essential: \emph{A dataset should include the most probable attack vectors starting from the attack entry point to the target node.}

\subsection{Attack Attributes}
Every attack type has a set of associated attributes. A DoS attack attribute could be the data transfer rate to the target node. An integrity attack attribute could be the injected signal characteristics. Every node has a unique reaction to different attribute values. For example, a DoS attack with moderate traffic rate may have insignificant impact on a controller that runs an RTOS that prioritizes control tasks over communication tasks. Therefore, for the published datasets to be useful for the wider research community, a good coverage for different attacks and their attributes is required. One challenge is that to produce a dataset with wider coverage, the number of attacks to be injected may grow exponentially, and techniques to limit this number may be needed, such as node similarity, injected signal impact, and discretization of continuous-value attributes. Automating \emph{relevant} attack identification from a given CPS is an important research topic that received little attention from the research community.

\vspace{2mm}\noindent $\diamond$ Essential: \emph{A dataset should have sufficient coverage for attack types and attack attributes as related to the underlying CPS.}

\section{Dataset Quality} \label{sec:DATASET-FEATURES}
In this section, we discuss required qualities of a CPS dataset for more-efficient learning. Most discussed aspects are projections of the classical dataset quality requirements for any machine learning domain.

\vspace{1mm}\noindent\emph{Data Labeling}. Multiple dataset records could belong to a single attack. Hence, each attack should have a unique identifier used as a label for each data record that belongs to the attack, in addition to the classical normal/attack or multiclass labeling. This allows for the development of more sophisticated attack detection and prevention architectures that correlate data rather than work on individual records. To add unique attack ID labels, a degree of automation is required in monitoring and collecting the dataset. Manual labeling in this case would be a daunting task. In addition, each dataset record has to be labeled with the associated mode of operation of the system, whether a normal mode of operation or a failure mode, as discussed in section \ref{subsec:systems-modes}. Therefore, a dataset label is a pair $\langle a,m \rangle$, where $a \in A$ is the attack unique ID and $m \in M$ is the associated system mode. 

\vspace{1mm}\noindent\emph{Class Imbalance}  \label{subsec:class-imbalance}. Class imbalance is typical for CPS datasets. Besides undersampling and oversampling techniques \cite{he2009learning}, the problem could be addressed at the dataset generation time by injecting more attacks. This is feasible only if the attack injection is partially or fully automated, which requires executable attacker models that could be used in penetration testing. Attacker models have recently gained research interest due to the increasing demand for automated penetration testing and big datasets \cite{deloglos2020attacker}.

\vspace{1mm}\noindent\emph{Data Redundancy}. Data records for the same CPS state do not carry new information. A dataset reporting one week of normal operation with a 1-sec time resolution could be reduced to an equivalent few-hours dataset. Redundant data consume more training resources when developing detection and response systems without a real benefit. Elimination of redundant data is crucial.

\vspace{1mm}\noindent\emph{Time Synchronization}. With synchronized nodes, the timestamps of dataset records could be used to reconstruct the causal relationship between system and attack events. Out-of-sync nodes introduce out-of-sequence data records, resulting in incorrect models of the system behavior. This is particularly important for synthesized datasets formed  by either merging physical and cyber data or by merging malicious network traffic with normal traffic.


\vspace{1mm}\noindent\emph{Dataset Scale}. Existing datasets are generated from small-scale CPSs. Real world systems may have thousands of sensors, actuators, and controllers connected using complex network architectures. Data generated from real systems is different from testbed-generated data in magnitude and correlations between variables. It is difficult to guarantee that solutions developed using testbed datasets would be scalable. Unfortunately, this is an inherent problem with testbeds that is unlikely to be solved in the near future given the scarcity of real world CPS security datasets. We briefly discuss one potential solution using virtual testbeds in Section \ref{subsec:virtual-testbeds}.

\begin{figure}[tb!]
	\centering
	\includegraphics[scale=0.15]{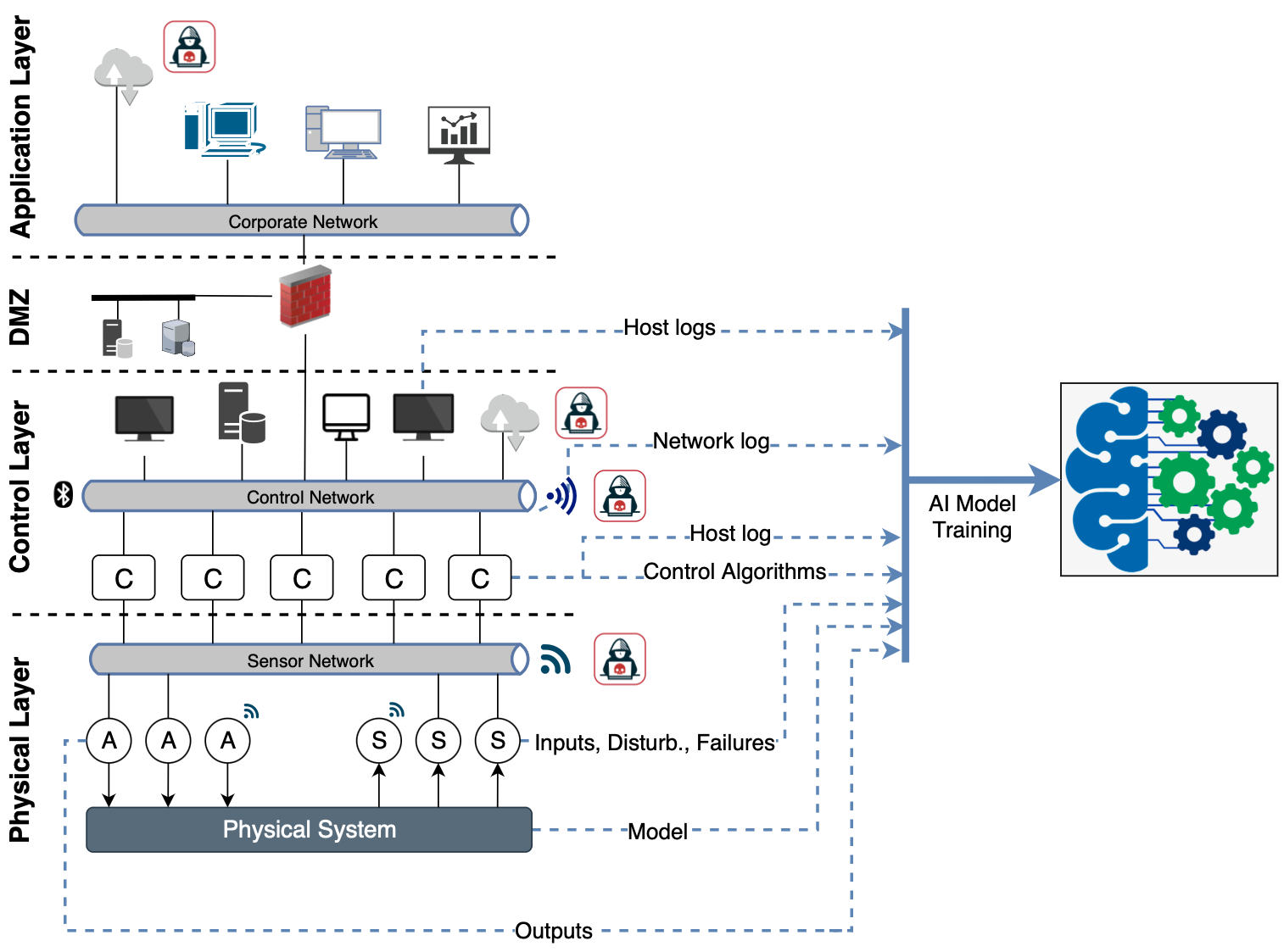}
	\caption{Dataset elements discussed in the paper. Attack diversity and dataset attributes are not explicitly depicted.}
	\label{fig:CPS-ARCH-ELEMENTS}
\end{figure}

\vspace{1mm}\noindent Figure \ref{fig:CPS-ARCH-ELEMENTS} summarizes the dataset elements as extracted from the general CPS architecture presented in Figure \ref{fig:CPS-ARCH}.

\section{Testbeds and Scalability} \label{sec:TESTBEDS}
Available datasets lack some or most of the features presented in this paper. As a result, some research groups opted for building their own physical or virtual testbeds to have full control on the experiments. In principle, the key benefits of the testbeds for the research community are twofold: \begin{inparaenum}[(1)] \item generate public datasets, and \item facilitate the replication of the testbed in less time and/or with less cost. \end{inparaenum} Unfortunately, only few of these testbeds are used to generate public datasets, as summarized in Section \ref{sec:related-work}. Furthermore, to the best of author's knowledge, none of the available testbeds has detailed documentation to enable other research teams to replicate the testbed. The result is a forest of isolated testbeds with local benefits only to the individual research groups. For a comprehensive survey on available testbeds, the reader is referred to \cite{Conti2021}.

\subsection{Virtual Testbeds} \label{subsec:virtual-testbeds}
Scalability is a major issue with physical testbeds, as it is not possible to match the scale of real systems with limited time and budget. The resulting dataset will remain small-scale when compared to real systems. Virtual testbeds represent a viable alternative. The key advantages of virtual testbeds when compared to physical testbeds are the minimum development time, scalability, and adaptability. It is possible to add hundreds of virtual nodes programatically to an existing virtual system using a set of pre-configured images, and both physical system and network simulation tools are available.  Also, virtual testbeds can adopt easily by modifying cyber node images or replacing the physical system simulator to support a different CPS application. However, one challenge is the availability of virtual images for industrial nodes such as sensors, actuators, controllers, PLCs, and embedded systems in general, compared to virtual images for IT components such as PCs, switches, and routers. Although virtualization may not be as accurate as real system behavior, high-fidelity models for both physical and cyber components can well approximate the real-world behavior.

\subsection{Dataset Generation}
The generation of CPS datasets is a laborious process. Despite the advancement of penetration testing tools that automate many of the attacker's actions, the overall process is still human-based. This results in a slow release of different versions of the datasets. Full automation of the process, including human-CPS interaction, disturbance injection, attack design and injection, and automated data logging and labeling, is still a distant goal. 

\section{Discussion and Future Directions} \label{sec:DISCUSSION}
CPS datasets are domain and application-specific. Although a general architecture could be proposed, akin to Figure \ref{fig:CPS-ARCH}, the structure and content of data vary. Physical system data depends entirely on the dynamics of the system. An autonomous vehicle has different dynamics than a manufacturing plant. Cyber system data depends on the communication protocols used and the amount of human-CPS interaction. An autonomous drone that has minimum to no human intervention would have different cyber data than a manufacturing plant that runs in a semi-autonomous mode with frequent human intervention, even if the communication protocols happened to be the same. This fact makes the availability of a common benchmark dataset a very difficult goal to achieve.

The amount of data resulting from the elements proposed in this work may look excessive. Given the autonomous nature of CPSs, the collection of such data should not be problematic. Most embedded and RTOSs support multilevel host logging. Most networking equipment support mirroring, and if not, host networking data could be collected and aggregated. Sensor and actuator data are available either in the host controller or in the smart sensor and actuator embedded system. The true challenge is the analysis and extraction of meaningful patterns from such big data. As discusses earlier, CPS data is mostly redundant, with varying degrees depending on the domain. An essential data pre-processing task is to remove this redundancy before applying any analysis or AI learning algorithms. The automation of CPS data redundancy removal is an important research direction.

The injection of cyber attacks to generate a somewhat balanced dataset is a laborious task. Automated penetration testing can help, but the planning phase where the system is studied and vulnerabilities are selected for exploitation are mostly human-centric. Automated extraction and deployment of relevant critical attacks for a given CPS is an important, yet difficult, research goal.

The construction of physical testbeds to generate datasets proved to be of limited benefits to the research community. Virtual testbed development is a promising direction. Virtualization in the domain of embedded and CPSs is not yet a mature technology when compared to IT systems. Embedded systems are time-critical, and virtualization of embedded nodes needs to capture this time resolution on a coarse level as dictated by the application.

\section{Related Work} \label{sec:related-work}
This section is a survey on existing CPS security datasets. Table \ref{tab:Datasets} summarizes the elements of each dataset against the proposed elements in this paper. It should be noted that the literature has many papers describing laboratory testbeds for the purpose of conducting research on CPS security. However, not all datasets are released for public use. Also, there may be small-scale datasets not mentioned in Table \ref{tab:Datasets}. These datasets are not used in the literature and their features are already captured by other datasets in Table \ref{tab:Datasets}, so expanding the list will not contribute to the discussion. The interested reader is referred to the recent survey on datasets and testbeds in \cite{Conti2021}.

\subsection{IT Domain Datasets}
The most common datasets used to evaluate network based IDS are DARPA dataset \cite{Lippmann2000}, DARPA Operationally Transparent Cyber (OpTc) dataset \cite{DARPAOpTC2019,DARPAOpTC-Anjum2021}, KDDCup99 \cite{Bay2000}, NSL-KDD \cite{Tavallaee2009}, UNSW-NB15 \cite{Moustafa2015}, botnet dataset \cite{Garcia2014}, and CICIDS2017 \cite{Sharafaldin2018}. These datasets are not suitable for CPS security research because: \begin{inparaenum}[(1)] \item the collected traffic data represents generic IT networks, which lacks industrial communication protocols as well as the industrial traffic patterns, and \item no physical system is associated with the cyber system, hence no physical data is available, which represents a key distinguishing feature of CPS security.\end{inparaenum} 

\subsection{CPS Domain - Physical and Cyber Datasets}
The most widely-used datasets are generated by iTrust research center and maintained at iTrust website \cite{iTrustdataset}. These datasets contain network traffic data and sensor/actuator data. The Secure Water Treatment (SWaT) dataset is generated by a small-scale water treatment testbed with 51 sensors and actuators \cite{Goh2017, Mathur2016a}. The dataset includes 11 days of continuous operation, 7 days of normal operation and 4 days with attack scenarios. The SWaT testbed was used to generate two additional datasets (S317, CISS2019.A1) in the context of Critical Infrastructure Security Showdown (CISS/STUD), where a number of red teams are allowed to design and launch attacks in real-time on the testbed. The Electric Power and Intelligent Control (EPIC) dataset is generated from a small-scale electric power generation and distribution testbed and includes 30 minutes with 8 operating scenarios \cite{adepu2018epic,ahmed2020comprehensive}. A derived dataset from EPIC, Blaq\_0, was generated in a hackathon competition and contains mainly network pcap files. Finally, network traffic data for 40 IoT honeypots with public IP addresses was collected for 1.5 years to generate an IoT dataset \cite{LinAung2020}.

Researchers in collaboration with Oak Ridge National Laboratory (ORNL) developed three Industrial Control System (ICS) datasets. Dataset 1 is generated from a simple power generation testbed with a simulated physical process. The dataset includes natural, faulty, maintenance, and attack scenarios. Network data includes IDS alerts only \cite{Pan2015}. Dataset 2 is generated from a small-scale gas pipeline testbed. Network traffic is a stripped-out version of Modbus fields, so some features such as inter-arrival time, number of packets, and TCP traffic pattern are missing \cite{Beaver2013}. Dataset 3 is generated from a gas pipeline ICS, and was later found to be flawed with unintended patterns that led ML algorithms to identify attacks with 100\% accuracy. A new version of the dataset with more randomness was released \cite{Turnipseed2015new}. The three testbeds and datasets are described in \cite{Morris2011a,morris2014industrial}, and maintained at the website \cite{MorrisaWebLink}. Researchers at Queensland University of Technology, Australia, developed a small-scale ICS testbed and generated two datasets for two different industrial processes \cite{Myers2018}. Physical process measurements were extracted from control device logs, and network traffic was captured in pcap files. One of the datasets, QUT\_S7, is available for download at \cite{Myers-Dataset}. The authors in \cite{lin2021ricsel21} generate a dataset called RICSel21 from a virtual testbed for a power network. The captured packets are for IEC-60870-5-104 protocol. 

\subsection{CPS Domain - Cyber Datasets}
A number of datasets that contain industrial network traffic data only have been published. A synthesized dataset for IEC 61580 substations has been developed to cover GOOSE messaging, a Manufacturing Message Specification (MMS) that is  prevalent for automated protection and control in modernized substations \cite{Biswas2019, IEC61580-GOOSE}. The substation measurements are provided separately in CSV files for the normal scenario only. The authors assume physical data is the same under attack scenarios, given the open-loop nature of the system. This assumption is not very accurate because the state estimator node, and potentially the IDS node, can see only the manipulated data stream, and not the true data. ELECTRA dataset has been generated to model the behavior of the control system of an electric traction substation used in a real high-speed railway area \cite{Gomez2019,ELECTRADS2019}. The dataset supports Modbus and S7Comm protocols. Lemay is a SCADA network dataset for Modbus/TCP protocol for a small electrical network \cite{Lemay2016,LemayDataset}. Several other \emph{raw} datasets have been collected in the form of pcap files as a representative of CPS network traffic, without labeling or feature extraction. 4SCIS Geek Lounge dataset is collected from the geek lounge of 4SCIS annual summit and contains network traffic data from a variety of industrial SCADA and PLC equipment, RTUs, servers, and industrial network equipment \cite{4SCISDataset}. The dataset comprises multiple pcap files with Modbus/TCP, BACnet, Ethernet/IP, and CIP protocols. S4X15CTF dataset was collected from Security Scientific Symposium 2015 (S4x15) during the Capture-the-Flag (CTF) competition using a variety of PLC's, HMI workstations and standard corporate IT PC's \cite{S4X15CTFDataset}. DEFCON23 dataset is collected from ICS village at DEFCON conference with hardware from different vendors supporting Modbus TCP, Profinet DCP, Profinet IO, and Profinet DCP protocols \cite{DEFCON23}. For all these datasets, since the system lounge is open to event participants to experiment with, the data may or may not include cyber attacks. With the lack of attack labels, it is challenging to use these datasets for research work.

\subsection{CPS Domain - Physical Datasets}
Water Distribution dataset (WADI) was generated by iTrust from a testbed having 123 sensors and actuators. The dataset includes physical measurements for 16 days of continuous operation, including 2 days with attack scenarios \cite{MujeebAhmed2017}. The WADI testbed was used to generate two additional datasets to support the BATtle of Attack Detection ALgorithms (BATADAL) competition. One dataset is for one year with no attacks, and the other dataset is for 6 months and includes several attacks with labels \cite{Taormina2018battle}. The Hardware In the Loop (HIL) Augmented ICS Security (HAI) dataset is developed using a laboratory testbed for power generation and storage \cite{shin2020hai}. The dataset contains physical plant measurements with different injection attacks on set points, process variables, control outputs, and control parameters. Attacks are automatically-generated and some attacks are stealthy. A system architecture to expand the dataset to include network and host data and more attack scenarios based on MITRE ATT\&CK framework is reported in \cite{Choi2020}. 

Table \ref{tab:Datasets} highlights the limitations of existing datasets. Host logs are entirely missing. System failures are not considered in almost all datasets. There is no diversity in injected cyber attacks. DoS attacks are not considered in most datasets. Attack labeling is mainly for individual records and lacks the complete attack vector perspective. Class imbalance is a true issue due to the difficulty of administering cyber attacks that are mostly injected manually. Finally, an issue that is not clear from Table \ref{tab:Datasets} is the difficulty of extracting information from published datasets. Constructing Table \ref{tab:Datasets} has taken a great deal of effort with yet some inaccuracies due to the poor documentation of most published datasets. Lack of proper documentation, besides dataset content, compel researchers to generate their own datasets, resulting in duplicated efforts with no reward to the research community.

\section{Conclusion} \label{sec:CONCLUSION}
Existing CPS security datasets lack some essential features that are necessary to build robust AI solutions. The dataset elements presented in this paper have the potential to enable the development of high-performance AI models and to facilitate dataset reuse. Physical testbeds, although useful to gain initial insights, are not a scalable solution and are unlikely to play a major role in the future of CPS security research. Virtual testbeds have the potential to overcome the problems of dataset scalability, high cost and long development time of physical testbeds, as well as testbed reuse.

Several research challenges remain unsolved. First, for a given CPS, the identification of the most-important attack scenarios as well as the automated design of such attacks is largely unsolved. Most of existing research work dealing with threat modeling focuses on the design of attacks on the physical level rather than a complete cyber-physical scenario. Practically, the process largely depends on the penetration tester's experience, and it is hard to find a joint expertise in both the physical and cyber domains. Second, virtualization progress in the embedded systems field is slower than its IT counterpart. Without a complete virtualization echo system, fast and efficient development of near real-time virtual testbeds is a challenging goal. Third, the domain of CPS security is still in its infancy, lacking any standards or benchmarks. The development of benchmarks is essential to compare developed models akin to other domains of AI.






\ifCLASSOPTIONcaptionsoff
  \newpage
\fi

\bibliographystyle{ieeetranurldate}
\bibliography{library}

\onecolumn
\begin{sidewaystable}[!htbp]
	\centering
	\begin{tabular}{lccccccccccccccc}
		\toprule
		Dataset &  \multicolumn{4}{c}{Physical System Data} & \multicolumn{2}{c}{Cyber System Data} & \multicolumn{3}{c}{Cyber Attacks} & \multicolumn{4}{c}{Dataset Features} \\
		{} & I/Os & Failures & Model & Control Alg. & Network & Host & Integrity & DoS & Vectors & Label & Time Stamp & \% Attack & Size \\
		\midrule
		SWaT \cite{Goh2017} & \CIRCLE & - & - & - & \CIRCLE & - & \CIRCLE & - & - & \LEFTcircle \footnote{The labels for network packets are not included in the available dataset, but could be added knowing attack start and end times. It was not possible to locate this data for SWaT dataset, but it is available for WADI dataset} & Y & 6\% & (950K, 388M)  \\
		EPIC \cite{ahmed2020comprehensive} & \CIRCLE & - & - & - & \CIRCLE & - & - & - & - & - & Y & 0\% & (15K, 3.6M) \\
		ORNL DS-1  \cite{Pan2015} & \CIRCLE & - & - & - & \LEFTcircle \footnote{Snort alerts only.} & - & \CIRCLE & - & - & \LEFTcircle & N & 70\% & 160 K \\
		ORNL DS-2 \cite{Beaver2013} & \CIRCLE & - & - & - & \LEFTcircle \footnote{Modbus frames breakout. \label{modbus}}  & - & \CIRCLE & - & - & \LEFTcircle & N \footnote{\label{TimeStamp} Time interval between Modbus command and response packets} & 4\% & 40 K \\
		ICS Water DS-3 \cite{Turnipseed2015new} & \CIRCLE & - & - & - & \LEFTcircle \footref{modbus}  & - & \CIRCLE & \CIRCLE & \LEFTcircle \footnote{Reconnaissance attacks. \label{Reconn.}} & \LEFTcircle & N \footref{TimeStamp} & 27\% & 236 K \\
		ICS Gas DS-3 \cite{Turnipseed2015new} & \CIRCLE & - & - & - & \LEFTcircle \footref{modbus}  & - & \CIRCLE & \CIRCLE & \LEFTcircle \footref{Reconn.}  & \LEFTcircle & N \footref{TimeStamp}& 37\% & 98 K \\
		QUT\_S7 \cite{Myers-Dataset} & \CIRCLE & - & - & - & \CIRCLE & - & \CIRCLE & \CIRCLE & - & \LEFTCIRCLE \footnote{No labels for network data. \label{NetData}} & Y & 0.006\% & 5M \\
		RICSel21 \cite{lin2021ricsel21} & \CIRCLE & - & - & - & \CIRCLE  & - & \CIRCLE & \CIRCLE & -  & \LEFTcircle & Y & 84\% & 10 K \\
		\arrayrulecolor{gray}\hline \\
		IEC-61580 \cite{IEC61580-GOOSE} & \LEFTcircle \footnote{IED data for the normal scenario only, assuming same data for the attack scenarios (open loop system).} & \CIRCLE & - & - & \CIRCLE  & - & \CIRCLE & \CIRCLE & \LEFTcircle \footnote{Network traffic is grouped according to the attack type. Feature extraction and attack labeling are needed. \label{labelling}}  & \LEFTcircle \footref{labelling}  & Y & 62\% & 430K \\
		ELECTRA Modbus \cite{Gomez2019} & -\footnote{Only the payload data, with no association with sensors/actuators. \label{payload}} & - & - & - & \CIRCLE  & - & \CIRCLE & - & - & \LEFTcircle  & Y & 15\% & 16M \\
		ELECTRA S7comm \cite{Gomez2019} & -\footref{payload} & - & - & - & \CIRCLE  & - & \CIRCLE & - & - & \LEFTcircle  & Y &   31\% & 387M  \\
		Lemay SCADA \cite{LemayDataset} & -\footref{payload} & - & - & - & \CIRCLE  & - & \CIRCLE & - & \LEFTcircle  & \LEFTcircle & N \footnote{Relative timing of subsequent packets only. \label{RelTim}} &   3.4\% & 912K  \\
		Lemay Covert \cite{LemayDataset} & -\footref{payload} & - & - & - & \CIRCLE  & - & - & - & \LEFTcircle  & \LEFTcircle & N\footref{RelTim} &   100\% & 1.6M  \\
		4SICS \cite{4SCISDataset} & - & - & - & - & \CIRCLE  & - & - & - & - & - & N\footref{RelTim}& UNK   & 3M  \\
		S4x15 \cite{S4X15CTFDataset} & - & - & - & - & \CIRCLE  & - & - & - & - & - & N\footref{RelTim} & UNK   & 310K  \\
		DEFCON23 \cite{DEFCON23} & - & - & - & - & \CIRCLE  & - & - & - & - & - & N\footref{RelTim} & UNK   & 1.4M  \\
		\arrayrulecolor{gray}\hline \\
		WADI  \cite{MujeebAhmed2017} & \CIRCLE & - & - & - & - & - & \CIRCLE & - & - & \LEFTcircle & Y & 1\% & 960K \\
		BATADAL \cite{Taormina2018battle} & \CIRCLE & - & - & - & - & - & \CIRCLE & - & - & \LEFTcircle & Y & 4\% & 13K \\
		HAI \cite{shin2020hai} & \CIRCLE & - & - & - & -  & - & \CIRCLE & - & - & - & Y & 1.8\% & 1M  \\
		UCI-Water \cite{UCIRVDataset} & \CIRCLE & \CIRCLE & - & - & -  & - & - & - & - & - & N & - & 527  \\
				\bottomrule
	\end{tabular}
	\caption{Elements of published datasets as compared to the proposed elements. Attack vector column refers to the full representation of the attack vector in the dataset, starting from the attack entry point. Dataset label column refers to the labeling of all the records of the same attack scenario using unique identifiers. Half-filled circles refer to individual record labeling, whether binary or multiclass. \% Attacks column indicates the percentage of attack records with reference to the whole dataset size, and represents the class imbalance. Size column is an estimation of the number of records in the physical dataset, number of packets in the cyber dataset, or both (physical, cyber), as per the dataset type. Large number of DoS packets are excluded from the estimation. Disturbances element was omitted as it was not possible to obtain the underlying CPS model for any of the published datasets to investigate whether disturbance variables are included in the physical dataset. UNK: Unknown.}
	\label{tab:Datasets} 
\end{sidewaystable}

\end{document}